\documentclass[prl,twocolumn]{revtex4}
\usepackage{amssymb}

\draft
\input{tcilatex}

\begin{document}

\title{Suppression of Quantum Phase Interference in Molecular Magnets Fe$%
_{8} $ with Dipolar-Dipolar Interaction}
\author{Zhi-De Chen$^{1}$, J.-Q. Liang$^{2}$, and Shun-Qing Shen$^{3.\ast }$}
\affiliation{$^{1}$Department of Physics and Institute of Modern Condensed Matter
Physics, Guangzhou University, Guangzhou 510405, China\\
$^{2}$Institute of Theoretical Physics and Department of Physics, Shanxi
University, Taiyuan, Shanxi 030006, China\\
$^{3}$Department of Physics, The University of Hong Kong, Pokfulam Road,
Hong Kong, China}

\begin{abstract}
Renormalized tunnel splitting with a finite distribution in the biaxial spin
model for molecular magnets is obtained by taking into account the dipolar
interaction of enviromental spins. Oscillation of the resonant tunnel
splitting with a transverse magnetic field along the hard axis is smeared by
the finite distribution which subsequently affects the quantum steps of
hysteresis curve evaluated in terms of the modified Landau-Zener model of
spin flipping induced by the sweeping field. We conclude that the
dipolar-dipolar interaction drives decoherence of quantum tunnelling in
molcular magnets Fe$_8$, which explains why the quenching points of tunnel
spliting between odd and even resonant tunnelling predcited theoretically
were not observed experimentally.
\end{abstract}

\maketitle

Macroscopic quantum phenomena in magnetic molecular clusters have been being
an attractive field in recent years\cite%
{Gunther94,Chudnovsky98,Wernsdorfer99,Garg93,Liang00,Chen02,Chudnovsky01,Leuenberger00}%
. Octanuclear iron(III) oxo-hydroxo cluster Fe$_{8}$ is of special interest
because it shows not only regular steps in hysteresis curve but also
oscillation of the tunnel splitting due to the quantum phase interference
[3]. Oscillation of tunnel splitting of the ground state with respect to the
external field along the hard axis was predicted theoretically by Garg [4]
as a consequence of the quantum phase interference of tunnel paths, and it
was subsequently generalized to tunnelling at excited states and resonant
tunnelling for quantum transition between different quantum states with its
x component of the spin $S_{x}=-10$ and $10-n$ (along the easy axis of Fe$%
_{8}$) recently [6]. The quenching points between even and odd $n$ have a
shift $\pi /2$. However, a serious problem, why the theoretically predicted
shift of quenching points of tunnel splitting between odd and even resonant
tunnelling was not observed in the experimental hysteresis curves [3,6,9],
remains to be solved. This is the main motivation of this Letter. Here we
use the Landau-Zener model [3,8,9,10] to describe the spin flipping induced
by the sweeping field with a modified bare tunnel splitting considering the
dipolar interaction with environmental spins. There are two basic
interactions to be considered: spin-phonon and spin-spin interactions. For
the molecular magnets Fe$_{8}$ in mK temperature region, the spin-phonon
interaction [8] can be safely ignored as the spin-lattice relaxation time is
extremely long [11]. The interaction between the big spin and the
environmental spins was considered as the main source of decoherence of
tunnelling in magnetic macromolecules [12] and recently it was shown that
the nuclear spin plays an important role in magnetic relaxation [13,14]. In
this Letter, starting from the mean field approximation, the dipolar
interaction is treated as a local stray field $\vec{h}$ (see the following)
with a Gaussian distribution. The tunnel splitting in the Landau-Zener
transition rate should be considered as an average over the local stray
field $\vec{h}$. In doing so we find that the quenching of the tunnelling
due to quantum interference is suppressed by the local stray field, and the
steps in the hysteresis curve corresponding to odd resonant tunnelling are
understood.

We start with the biaxial spin model for the molecular magnets Fe$_{8}$
[3-5]. The Hamiltonian is given by [15] 
$$
H=K_{1}S_{z}^{2}+K_{2}S_{y}^{2}-g\mu _{B}\mathbf{S}\cdot (\mathbf{B}+\vec{h}%
),\eqno(1) 
$$%
where $K_{1}>K_{2}>0$ and $\mathbf{B}$ is the external magnetic field. The
term $-g\mu _{B}\mathbf{S}\cdot \vec{h}$ is the dipolar-dipolar interaction
between the magnetic molecular cluster and the environmental spins, i.e. $%
\vec{h}=\sum_{j}J_{ij}\mathbf{S}_{j}$, where the summation runs over the
neighboring clusters. Strictly speaking, this should be a many-body problem.
In this Letter, $\vec{h}$ is treated approximately as a local stray field, $%
\vec{h}=\sum_{j}J_{ij}\left\langle \mathbf{S}_{j}\right\rangle $. Both
experimental [14] and the Monte Carlo study[16,17] show that $\vec{h}$ has a
random distribution with a distribution width in proportion to $(1-|M|)$ and
its mean value proportional to $M$ where $M$ is the total magnetization of
the system. Here we assume that $\vec{h}$ has a Gaussian distribution with
an equal distribution width in all directions [18] 
$$
P(\vec{h})=\frac{1}{(2\pi \sigma ^{2})^{3/2}}\exp \left[ -(\vec{h}-\vec{h}%
_{0})^{2}/2\sigma ^{2}\right] .\eqno(2) 
$$%
To simulate the experimental setup [3], the external magnetic field is taken
to be$\ \mathbf{B}=\{B_{x},0,B_{z}\}$: a uniform field $B_{z}$ along the
hard axis and the sweeping field $B_{x}$ on the easy axis $B_{x}=n\Delta
B\pm ct$ where $n$ is integer, $\Delta B$ is the field interval between
neighboring resonant tunnelling and $c=dB_{x}/dt$. In the following
calculation, we take $K_{1}=0.310$K, $K_{2}=0.229$K, and $c=0.1$T/sec for
the molecular magnets Fe$_{8}$ [3].

Theoretically, quantum tunnelling for a spin system without the local stray
field can be understood in the instanton method [4-6], the Landau-Zener
model [8-10], and by diagonalizing the Hamiltonian numerically [3, 19]. The
instanton method can give the tunnel splitting. When the field along the
easy axis satisfies the resonant condition, $B_{x}+h_{x}=n\Delta B,$ the
transition rate, while the field on the easy axis sweeps over the resonant
point, is given by the Landau-Zener transition formula [8-10], $%
P_{LZ}=1-\exp \left( -\pi \Delta _{n}^{2}/\nu _{n}\right) ,$ where $\Delta
_{n}$ is the tunnel splitting and $\nu _{n}=2g\mu _{B}\hbar (2s-n)c$. It
should be noted that in this way we have assumed tacitly that the tunnel
splitting for all the spins inside the resonant window are the \textit{same}
and thus all the spins tunnel with the \textit{same} transition rate.
However, when the local stray field due to the dipolar-dipolar interaction
is taken into account, such a picture should be modified. A random
distribution of local stray fields like Eq.(2) with a distribution width $%
\sigma \sim 0.05$T typical for the molecular magnets Fe$_{8}$ will block the
resonant tunnelling of either the ground state or the low-lying excited
states [13]. Nevertheless, such a problem can be circumvented by using the
sweeping field along the easy field. When $B_{x}$ sweeps over the resonant
point, it will make the spins with different $h_{x}$'s to satisfy the
resonant condition, and allows continuous relaxation. Since the tunnel
splitting is very sensitive to the transverse local fields $B_{z}+h_{z}$,
and $h_{y}$ [4-6], the spins tunnel with \textit{different} tunnel splitting
while $B_{x}$ is sweeping over the resonant point. Consequently, the spin
transition rate observed in the experiment should be given by 
$$
\left\langle P_{LZ}\right\rangle \simeq 1-\exp \{-\pi \left\langle \Delta
_{n}^{2}\right\rangle /\nu _{n}\},\eqno(3) 
$$%
where $\left\langle \cdots \right\rangle $ represents the average over the
distribution of the local stray field, i.e., $\left\langle \Delta
_{n}^{2}\right\rangle =\int \Delta _{n}^{2}(\vec{h})p(\vec{h})d\vec{h}.$
Accordingly, the tunnel splitting extracted from the measured transition
rate should be $\sqrt{\left\langle \Delta _{n}^{2}\right\rangle }$ but not $%
\Delta _{n}$. In other words, the starting point to understand the
experimental observation should be $\sqrt{\left\langle \Delta
_{n}^{2}\right\rangle }$ instead of $\Delta _{n}.$ The two quantities are
qualitatively different from each other as we shall show in the following.

The instanton method [4-6] is efficient and powerful to evaluate the tunnel
splitting $\Delta _{n}$. The Lagrangian for the biaxial model Eq.(1) is 
$$
L(\mathbf{n})=-s\hbar (1-\cos \theta )\dot{\phi}-\left\langle \mathbf{n}%
\right\vert H\left\vert \mathbf{n}\right\rangle ,\eqno(4) 
$$%
where $\left\vert \mathbf{n}\right\rangle $ is the spin coherent state. With
the help of the mapping technique, $(\phi ,p=s\hslash \cos \theta )$ is
regarded as a pair of canonical variables. To calculate the excited state
tunnelling or resonant tunnelling, one needs to apply the Bohr-Sommerfeld
quantization rule $\doint pd\phi =n\hslash $ to define the classical orbits
(n is an integer). Then a propagator with both imaginary and real time will
be used to describe the tunnelling between two degenerate states,%
$$
\begin{array}{l}
K(n_{f},T/2;n_{i},-T/2)=\left\langle n_{f}\right\vert e^{iHT/\hbar
}\left\vert n_{i}\right\rangle \\ 
=\int d\Omega \exp \left[ \frac{i}{\hbar }\int_{-T/2}^{T/2}L(\mathbf{n})dt%
\right] .%
\end{array}%
\eqno(5) 
$$%
The tunnel splitting is found by integrating over two degenerate classical
orbits. In molecular magnets Fe$_{8}$ the local stray field is rather weak,
i.e. $g\mu _{B}|\vec{h}|/(K_{2}s)\ll 1$. We calculate the tunnel splitting
at the $n$th resonant tunnelling point with the transverse field $%
B_{z}+h_{z} $, and $h_{y}$, and obtain that, 
$$
\begin{array}{ll}
\Delta _{n} & \approx \frac{Q_{n}}{2}e^{-S_{c}^{n}}\left\{
|e^{2qh_{y}}+e^{-2qh_{y}}\right. \vspace*{0.3cm} \\ 
& +\left. 2\cos [2(s\pi -n\pi /2-d_{n}h_{z}-d_{n}B_{z})]|\right\} ^{1/2},%
\end{array}%
\eqno(6) 
$$%
where and $S_{c}^{n}$ is the instanton action 
$$
S_{c}^{n}=\int_{0}^{\pi -\phi _{n}}d\phi \sqrt{\frac{V(\phi )-E_{n}}{%
K_{1}(1-\lambda \sin ^{2}\phi )+(g\mu _{B}n\Delta h/s)\cos \phi }},\eqno(7) 
$$%
$$
\begin{array}{ll}
V(\phi )= & K_{2}s^{2}\sin ^{2}\phi -g\mu _{B}n\Delta hs\cos \phi \vspace*{%
0.3cm} \\ 
& -\frac{[g\mu _{B}(h_{z}+B_{z})]^{2}}{2K_{1}(1-\lambda \sin ^{2}\phi
)+2(g\mu _{B}n\Delta h/s)\cos \phi },%
\end{array}%
\eqno(8) 
$$%
$E_{n}$ is the energy of the $n$th excited state, $\phi _{n}$ is the turning
point determined by $V(\pi -\phi _{n})=E_{n}$, $Q_{n}$ is the pre-factor 
$$
Q_{n}\simeq 4\pi /\sqrt{V^{\prime \prime }(0)(2K_{1}+g\mu _{B}n\Delta h/s)},%
\eqno(9) 
$$%
$q=g\mu _{B}\pi \lambda ^{1/2}/2K_{2}(1-\lambda )^{1/2},$ $\lambda
=K_{2}/K_{1},$ and 
$$
d_{n}=\frac{g\mu _{B}}{2K_{1}}\int_{0}^{\pi }\frac{d\phi }{1-\sin ^{2}\phi -%
\frac{g\mu _{B}}{2K_{2}s}~n\Delta h\cos \phi }.\eqno(10) 
$$%
Using the parameters in Fe$_{8}$, it is found that the contribution from $%
h_{y}$ and $h_{z}$ to $S_{c}^{n}$ and thus $Q_{n}e^{-S_{c}^{n}}$ is very
small under the condition $g\mu _{B}|\vec{h}|/(K_{2}s)\ll 1$. The average
value of $\Delta _{n}^{2}$ is given by 
$$
\begin{array}{l}
\left\langle \Delta _{n}^{2}\right\rangle \approx \frac{Q_{n0}^{2}}{4}%
e^{-2S_{c0}^{n}}\left\{ e^{2q^{2}\sigma
^{2}}(e^{2qh_{0}}+e^{-2qh_{0}})\right. \vspace*{0.3cm} \\ 
+\left. 2e^{-2d_{n}^{2}\sigma ^{2}}\cos [2(s\pi -n\pi
/2-d_{n}h_{0}-d_{n}B_{z})]\right\} .%
\end{array}%
\eqno(11) 
$$%
where $Q_{n0}=Q_{n}(h_{z}=h_{y}=0),~S_{c0}^{n}=S_{c}^{n}(h_{z}=h_{y}=0)$. In
the absence of the stray field, i.e. $\sigma =h_{0}=0$, the above expression
reduces to 
$$
\begin{array}{c}
\left. \sqrt{\left\langle \Delta _{n}^{2}\right\rangle }\right\vert _{\sigma
=h_{0}=0}=\Delta _{n}(h_{x}=h_{y}=0) \\ 
=Q_{n0}e^{-S_{c0}^{n}}|\cos (s\pi -n\pi /2-d_{n}B_{z})|,%
\end{array}%
\eqno(12) 
$$%
which indicates the oscillation of the tunnel splitting with the transverse
field and a shift $\pi /2$ of quenching point between the odd and even
resonant tunnelling, recovering the results in the previous works [3-6].
This is known as a result of the quantum interference of the tunnelling
along two different paths. Qualitative difference between $\sqrt{%
\left\langle \Delta _{n}^{2}\right\rangle }$ and $\Delta _{n}(h_{x}=h_{y}=0)$
can now be seen by comparing Eq.(12) with Eq.(11). In the case of $B_{z}=0$
and integer spin, Eq.(12) predicts that odd n resonant tunnelling quenches
due to the quantum interference, while in the presence of the stray field
Eq.(11) gives non-zero tunnel splitting 
$$
\sqrt{\left\langle \Delta _{n}^{2}\right\rangle }_{q}\simeq \frac{\sqrt{2}}{2%
}Q_{n0}e^{-S_{c0}^{n}}\sqrt{e^{2q^{2}\sigma ^{2}}-e^{-2d_{n}^{2}\sigma ^{2}}}%
\eqno(13) 
$$%
for $h_{0}=0$. The quenching due to the quantum interference is suppressed
by the local stray field. In another word the quantum tunnelling for odd n
is decoherenced because of the dipolar interaction with the environmental
spins. The tunnel splitting of all six resonant tunnelling for the molecular
magnets Fe$_{8}$ with and without the local stray field are shown in Table
I. We see that $\sqrt{\left\langle \Delta _{n}^{2}\right\rangle }$ for an
odd n increases from zero while the random field become stronger. The random
field also increases the tunnel splitting of even resonant tunnelling. It
increases about $2.7$ times as $\sigma $ becomes as large as 0.08T, which
resolves the puzzling that the experimental observation is about $3.0$ times
larger than the numerical result for the tunnel splitting [3]. A detailed
evolution of the tunnel splitting with the distribution width around the
topological quenching points is shown in Fig.1. As the width of the
distribution is proportional to $(1-|M|)$, the calculated results for
different M are shown in Fig.1, which are in good agreement with the
experimental observation (see Fig.10 in Ref.[20]). One can see from Eq.(11)
that the main effect of $h_{0}$ is to provide an initial phase and thus
shifts the oscillation. For Fe$_{8}$, $h_{0}\simeq \sigma /4$ [16], and the
effect of modification for nonzero $h_{0}$ is almost omissible.

Table I: Tunnel splitting $\sqrt{\left\langle \Delta _{n}^{2}\right\rangle }$
(the unit is Kelvin) for Fe$_{8}$ in the case of $B_{z}=h_{0}=0$.

\begin{center}
\vspace*{0.2cm}%
\begin{tabular}{|l|l|l|l|l|}
\hline
$n$ & $\sigma =0.0$T & $\sigma =0.02$T & $\sigma =0.05$T & $\sigma =0.08$T
\\ \hline
$0$ & 8.399$\times $10$^{-10}$ & 8.547$\times $10$^{-10}$ & 1.087$\times $10$%
^{-9}$ & 2.312$\times $10$^{-9}$ \\ \hline
$1$ & 0.0 & 2.459$\times $10$^{-9}$ & 3.266$\times $10$^{-9}$ & 6.450$\times 
$10$^{-9}$ \\ \hline
$2$ & 3.414$\times $10$^{-8}$ & 3.473$\times $10$^{-8}$ & 4.418$\times $10$%
^{-8}$ & 9.393$\times $10$^{-8}$ \\ \hline
$3$ & 0.0 & 2.399$\times $10$^{-7}$ & 3.187$\times $10$^{-7}$ & 6.293$\times 
$10$^{-7}$ \\ \hline
$4$ & 2.015$\times $10$^{-6}$ & 2.050$\times $10$^{-6}$ & 2.608$\times $10$%
^{-6}$ & 5.544$\times $10$^{-6}$ \\ \hline
$5$ & 0.0 & 9.878$\times $10$^{-6}$ & 1.312$\times $10$^{-5}$ & 2.591$\times 
$10$^{-5}$ \\ \hline
$6$ & 6.224$\times $10$^{-5}$ & 6.333$\times $10$^{-5}$ & 8.055$\times $10$%
^{-5}$ & 1.713$\times $10$^{-4}$ \\ \hline
\end{tabular}
\end{center}

\FRAME{ftbpFU}{3.1652in}{2.1387in}{0pt}{\Qcb{Illustration of $\protect\sqrt{%
\left\langle \Delta _{n}^{2}\right\rangle }$ ($n=0,1$) around topological
quenching points due to quantum interference with different distribution
width $\protect\sigma $'s.}}{}{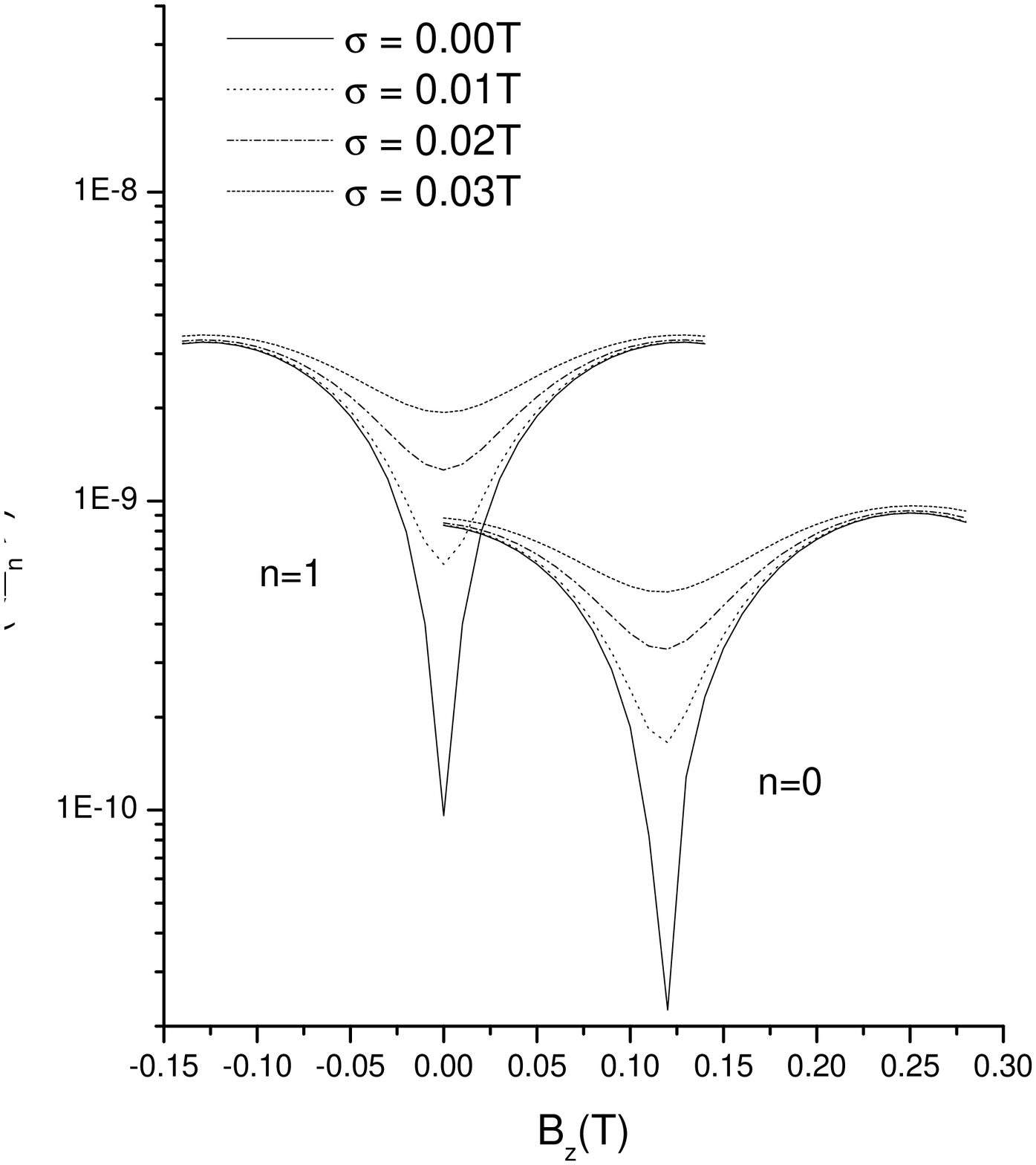}{\special{language "Scientific
Word";type "GRAPHIC";maintain-aspect-ratio TRUE;display "USEDEF";valid_file
"F";width 3.1652in;height 2.1387in;depth 0pt;original-width
4.5818in;original-height 3.0874in;cropleft "0";croptop "1";cropright
"1";cropbottom "0";filename 'fig1.eps';file-properties "XNPEU";}}

The oscillation of the tunnel splitting for $s=10$ with various distribution
width $\sigma $'s is shown in Fig.2. From Fig.2, it is shown that the
oscillation of the tunnel splitting due to quantum interference is
suppressed by the local stray field $\vec{h}$. For a distribution width $%
\sigma =0.05T$ which is estimated for Fe$_{8}$ [9,13], the oscillation of
tunnel splitting with respect to the field along the hard axis is still
visible, while the oscillation is suppressed completely for the width as
large as $0.08$T. In fact, when the distribution width approaches the half
oscillation period, the oscillation due to quantum interference disappears
and the classical behavior, i.e. tunnel splitting increases monotonously
with $B_{z}$, is resumed. The above analysis leads to a decoherence
mechanism for quantum interference due to the dipolar-dipolar interactions
between the spins without dissipation [12].

\FRAME{ftbpFU}{2.9577in}{2.0903in}{0pt}{\Qcb{ The oscillation of $\protect%
\sqrt{\left\langle \Delta _{0}^{2}\right\rangle }$ with different
distribution width $\protect\sigma $'s for s=10. From top to bottom: $%
\protect\sigma =0.08$T, 0.05T, 0.02T, and 0.0T.}}{}{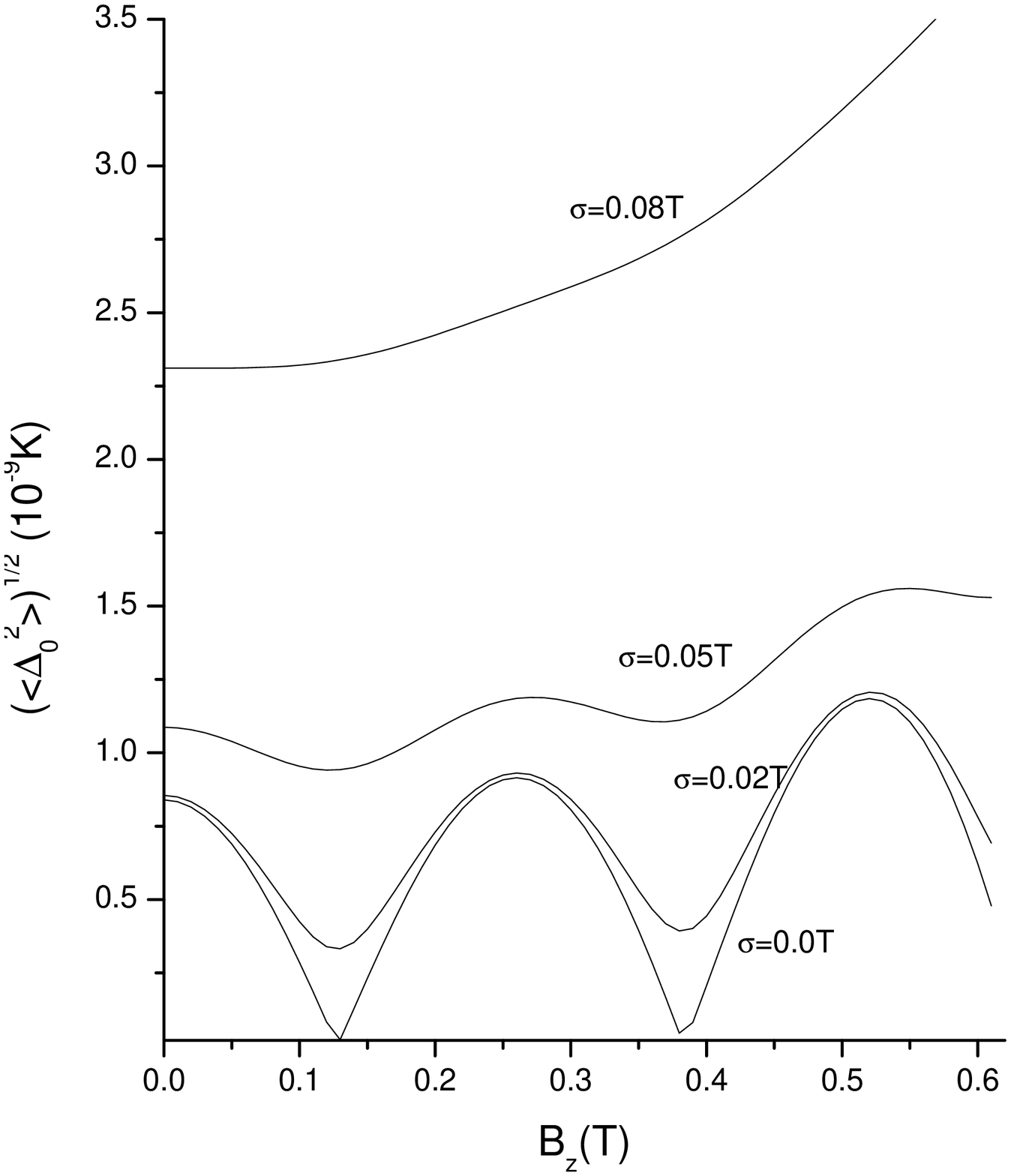}{\special%
{language "Scientific Word";type "GRAPHIC";display "USEDEF";valid_file
"F";width 2.9577in;height 2.0903in;depth 0pt;original-width
7.7721in;original-height 11.0757in;cropleft "0";croptop "1";cropright
"1";cropbottom "0";filename 'fig2.eps';file-properties "XNPEU";}}

The magnetization jump from the spin flipping at the resonant tunnelling can
be calculated from the modified Landau-Zener transition rate given in
Eq.(3). In principle the time evolution of the spin system in Eq.(1) can be
obtained by solving the time-dependent Schr\"{o}dinger equation $i\hbar 
\frac{\partial }{\partial t}\left\vert \Phi \right\rangle =H\left\vert \Phi
\right\rangle $, which contains a set of (2s+1) coupled differential
equations for the model in Eq. (1). It was shown [19] that the coupled
differential equations can be reduced to that of an effective two-level
system with the effective Hamiltonian. Here we have 
$$
H_{\mathrm{eff}}(t)=\left( 
\begin{array}{cc}
-(10-n)g\mu _{B}ct & \sqrt{\left\langle \Delta _{n}^{2}\right\rangle }/2%
\vspace*{0.3cm} \\ 
\sqrt{\left\langle \Delta _{n}^{2}\right\rangle }/2 & 10g\mu _{B}ct%
\end{array}%
\right) ,\eqno(14)
$$%
near the resonant condition and the time-dependent state is given by $%
\left\vert \Phi _{\mathrm{eff}}\right\rangle =a_{-10}(t)\left\vert
-10\right\rangle +a_{10-n}(t)\left\vert 10-n\right\rangle $. The tunnelling
splitting in Eq.(14) is $\sqrt{\left\langle \Delta _{n}^{2}\right\rangle }$
instead of $\Delta _{n}$ as we discussed. Correspondingly, the magnetization
jump from the $n$th resonant tunnelling is obtained as 
$$
\Delta M_{n}=\left\langle \Phi _{\mathrm{eff}}\right\vert S_{x}\left\vert
\Phi _{\mathrm{eff}}\right\rangle \left\vert _{t=+\infty }-\left\langle \Phi
_{\mathrm{eff}}\right\vert S_{x}\left\vert \Phi _{\mathrm{eff}}\right\rangle
\right\vert _{t=-\infty }.\eqno(15)
$$%
Numerical results are shown in Fig.3. It is worth emphasizing that the
resulting jump is modified as a rather smooth one due to the local stray
field. The hysteresis curves in Fig.3 are drawn with the initial condition
of $S_{x}=-10$, i.e.$,a_{-10}(t=-\infty )=1$. As it is shown in Fig. 3 the
steps in hysteresis curve are smeared gradually with increasing the
distribution width of the local stray field and the observed curve in
experiment [3,9] can be recovered from the present theory.\FRAME{ftbpFU}{%
2.9265in}{1.983in}{0pt}{\Qcb{Hysteresis curves with different distribution
width $\protect\sigma $'s.}}{}{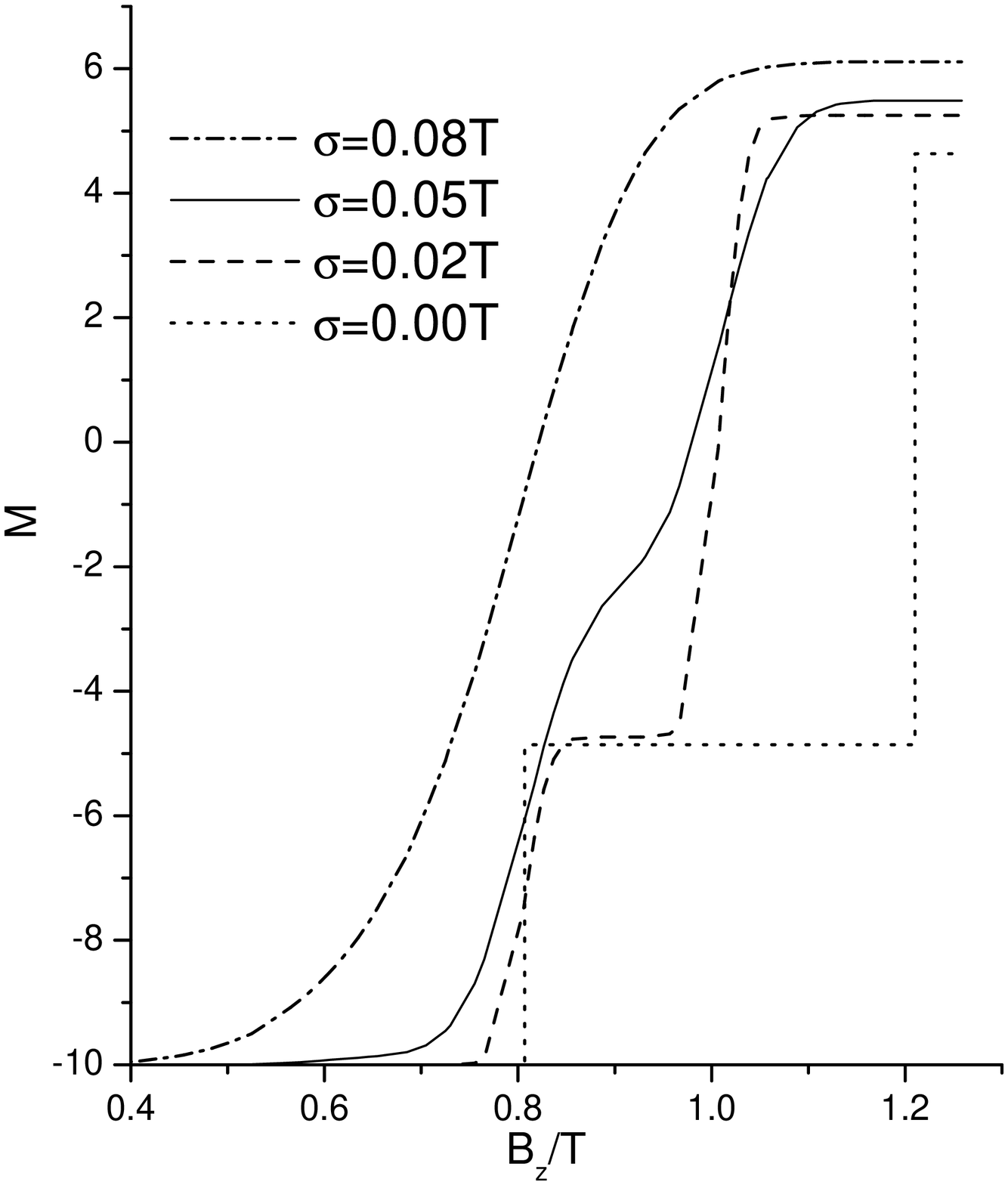}{\special{language "Scientific
Word";type "GRAPHIC";display "USEDEF";valid_file "F";width 2.9265in;height
1.983in;depth 0pt;original-width 7.7721in;original-height 11.0757in;cropleft
"0";croptop "1";cropright "1";cropbottom "0";filename
'fig3.eps';file-properties "XNPEU";}}

In this Letter, the local stray field is treated as a \textquotedblleft
frozen\textquotedblright\ one inside the resonant window. Strictly speaking,
both the width and the mean value of the distribution of the field should
vary with the time-dependent magnetization during the resonant tunnelling.
However, it should be noted that a \textquotedblleft
frozen\textquotedblright\ distribution is based on validity of the
Landau-Zener model. If the spins \textquotedblleft feel\textquotedblright\
the change of the local field due to spin flipping, the spin transition rate
is no longer the one in Eq.(3). In that case, one should consider the
non-linear Landau-Zener tunnelling [21] and the \textquotedblleft
hole-digging\textquotedblright\ mechanism [13,14]. This indicates that our
result is valid when the field sweeping rate is not too small such that the
evolution of the local field is relatively slower than the sweeping field.
Namely, the overlap time of two levels in resonance $\tau _{1}\sim \Delta
_{n}/(2\mu _{B}Sc)$ should be less than the characteristic relaxation time $%
\tau _{2}$ due to the dipolar-dipolar interaction. This means that the field
sweeping rate $c>c_{0}\sim \Delta _{n}/(2\mu _{B}Sc\tau _{2}).$ In Fe$_{8}$
[14,16] the ground state tunnelling $\Delta _{0}\sim 10^{-7}$K, $\tau
_{2}\sim 10^{-5}$sec, and c$_{0}$ is estimated to be $10^{-3}$T/sec., which
is in good agreement with the experimental condition [3,20]. On the other
hand, a finite distribution of tunnel splitting due to the local stray field
has a deeper impact on the magnetic relaxation. If all the spins tunnel with
the same tunnelling rate, the magnetic relaxation should obey the
exponential law, i.e. $e^{-\Gamma t}$ where $\Gamma =2P_{LZ}c/A$ where $A$
is amplitude of the ac field used in the experiment [20]. Instead, in the
present picture, there is a finite distribution of tunnel splitting $%
p(\Delta _{n})$ which will lead to a finite distribution of the relaxation
rate $p(\Gamma )$ characteristic of the complex system like spin glass [22].
Consequently the resulting relaxation will obviously deviate from the simple
exponential law as observed in the experiment [20]. Further analysis will be
provided elsewhere.

We have studied the effect of dipolar interaction between giant spins in the
molecular magnets Fe$_{8}$ in the mean field approximation which leads to a
Zimman term of the spin in the local stray field. Our main observation is
that the topological quench due to the quantum phase interference of tunnel
paths is suppressed by the finite distribution of the local stray field, and
the steps in the hysteresis curve corresponding to odd resonant tunnelling
are explained theoretically. Thus we conclude that the dipolar-dipolar
interaction leads to the decoherence of quantum tunnelling in Fe$_{8}.$
Finally it is worth pointing out that the mechanism of decoherence may not
be just limited in Fe$_{8}$, but can be generalized to other molecular
magnets such as Mn$_{12}$ since the local stray field due to the
dipolar-dipolar and hyperfine interactions always exists.

This work was supported by the Nature Science Foundation of China under
Grant No. 10075032, the Fund \textquotedblleft Nanomic Science and
Technology\textquotedblright\ of Chinese Academy of Science, the RGC grant
of Hong Kong, and the CRCG grant of the University of Hong Kong.

*Electronic address: sshen@hkucc.hku.hk

\end{document}